\documentclass{article}
\usepackage[utf8]{inputenc}
\usepackage{graphicx}
\usepackage{authblk}
\usepackage{caption}
\usepackage{subcaption}
\usepackage{changepage}
\usepackage{cite}
\usepackage{tcolorbox}
\usepackage[left=4cm,right=4cm]{geometry}

\title{``Stay Nearby or Get Checked'': \\ A Covid-19 Lockdown Exit Strategy \\[2ex]}
\author{Jan-Tino Brethouwer$^1$, Arnout van de Rijt$^{2,3,}$\thanks{corresponding author: arnoutvanderijt@gmail.com}, Roy Lindelauf$^4$, and Robbert Fokkink$^1$}
\date{%
    $^1$TU Delft, Delft Institute of Applied Mathematics \\%
    $^2$European University Institute, Political and Social Sciences\\%
    $^3$Utrecht University, Sociology\\%
    $^4$Netherlands Defence Academy, Faculty of Military Science, Intelligence and Security\\[4ex]%
    \today}

\begin{document}

\maketitle

\begin{abstract}
This paper repurposes the classic insight from network theory that long-distance connections drive disease propagation into a strategy for controlling a second wave of Covid-19. We simulate a scenario in which a lockdown is first imposed on a population and then partly lifted while long-range transmission is kept at a minimum. Simulated spreading patterns resemble contemporary distributions of Covid-19 across EU member states, German and Italian regions, and through New York City, providing some model validation. Results suggest that our proposed strategy may significantly flatten a second wave. We also find that post-lockdown flare-ups remain local longer, aiding geographical containment. Public policy may target long ties by heavily focusing medical testing and mobility tracking efforts on traffic and transport. This policy can be communicated to the general public as a simple and reasonable principle: Stay nearby or get checked.\end{abstract}

\section{Introduction}
Many countries facing the spread of Covid-19 are currently exiting a lockdown regime in which person-to-person contact was severely restricted. The constraints placed on social and economic interaction have high cost. How can we intelligently lift these constraints and open up society now that new infections have dwindled?

Here we explore the leverage gained from differentiating between short-distance and long-distance ties in post-lockdown policy. The idea is that the blockage of transmission through long-distance ties increases the effective \emph{diameter} of a network, which is inversely related to the speed of propagation \cite{kretzschmar2009mathematical, wallinga1999perspective}. In practice, such geographic differentiation may be achieved through location tracking technologies and prioritization of non-local travel and transport in policy restrictions, enforcement and medical testing. The relative sparsity of long-range ties may make tight control feasible through a focus of resources on a small number of key individuals or interactions.

Results show that reductions in transmission through long-range ties slow down Covid-19 to a much greater extent than reductions in short-range ties. Selective scrutiny of long-distance ties has two added benefits: Post-lockdown flare-ups of Covid-19 are local, allowing geographically focused interventions that are of limited economic damage and logistically more feasible. And social toll is diminished, as the intimacy of human relations and need for face-to-face contact are known to decrease sharply with geographical distance \cite{park1924concept, zipf1949human, 
granovetter1983strength, marsden1984measuring, 
latane1995distance, 
groh2013geographic, 
kaltenbrunner2012far}. 

Our paper is organized as follows. In section 2 we describe our proposed strategy. In section 3 we describe our mathematical model and validate it against the data. In section 4 we present the results of model simulations. Section 5 summarizes our findings and recommendations.

\section{Related work}

Social network models of disease spreading have been around for decades. What sets our work apart is an analysis of the epidemiological leverage of government policies that differentiate long-distance from short-distance ties in social networks.

\subsection{Social network models of infectious disease spread}

Many epidemiological studies assume random mixing of individuals within demographic subgroups (e.g. by age) \cite{manzo, PLRKED}. However, most contact occurs between people who live very close to one another \cite{butts2002spatial, ferguson2006strategies}. We draw on the well-known small-world model of Watts and Strogatz \cite{watts1998collective} to capture the fundamental difference in viral risk between close-range and long-range ties: Close-range ties connect infected individuals with others who are already infected or are about to regardless. Long-range ties expose faraway contacts who would otherwise not be at risk and who may in turn infect others who are otherwise safe.

The small-world approach to the study of epidemiological dynamics is not new. Network analysis was introduced into mainstream epidemiology at the turn of the century to explicitly incorporate the contact structure among individuals. It is well known that diffusion processes on networks depend on the corresponding connectivity patterns \cite{Kleinberg2007Cascading}. Research has shown that subtle features of network structure can have a major impact on the outcome of an epidemic \cite{SunGao2007Dynamical, Carnegie2018Effects}. Small-world networks are almost identical to lattice networks in which viruses spread slowly and locally \cite{watts1998collective, kleinberg2000navigation}. The subtle difference is a small portion of ties to distant localities on the lattice, producing a dramatic reduction in a network's diameter, which is inversely related to viral spread. The rapid spread of viruses through small-world networks makes them hard to contain in time within confined regions of a population. Past work has examined the effects of a range of network models on epidemiological dynamics \cite{keeling2005networks, Newman2002Spread,MooreNewman2000}.

A small-world SEIR model was used in~\cite{YBHB} to model an influenza outbreak in the city of Orhan. In a study closest to ours, Small and Tse investigate disease spread in a small-world network with separate infection probabilities for short-distance and long-distance ties \cite{SamllTse2005Small}. Using a SEIR model of the SARS epidemic dynamics they find that exponential growth in infection occurs upon onset of several non-local infections. They conclude that key to capturing the empirically observed transmission dynamics is differentiating local from non-local transmission probabilities. We build on this observation to explore the leverage that the targeting of Covid-19 post-lockdown policies at reductions of non-local transmission may provide to global, national, or regional policy makers. 

\subsection{Interventions}
In epidemiological models effects of both general and targeted interventions on disease spread have been studied \cite{Halloran2008Modeling}. General interventions such as social distancing and school closures aim to bring down overall infection probabilities or those within and between demographic subgroups \cite{PLRKED}. Targeted interventions seek to identify high-risk individuals: Antiviral treatment and household isolation of identified cases, prophylaxis and quarantine of household members. We propose a different kind of targeting that is not aimed at specific \emph{nodes} but at high-impact \emph{edges} of a network.

A challenge faced by contemporary policy makers is when and how to ease interventions. How can a second wave be minimized while at the same time preventing enormous economic costs? It is well known that when a lockdown is lifted, a virus tends to re-appear \cite{Ferguson2020}. Therefore it is of paramount importance to find ways to regain some form of normal life (alleviating lockdown) while at the same time preventing the virus from going viral again. The main idea proposed here is that restricting certain high-risk interactions within the social network may be a better strategy than to restrict those of an entire population. `Long-distance' ties represent interaction between individuals that are distant to each other in a network. Typically this means they are also physically distant, i.e. think of a truck driver that delivers goods to a company on the other side of the country or individuals traveling by plane that encounter each other at airports and in airplanes where social distancing is difficult or next to impossible. Small-world models suggest that long-distance ties greatly accelerate the speed of transmission. Long-range ties stemming from infected individuals allow disease to start spreading in distant other localities and much more often lead to not-yet-infected individuals and regions. At a global level, long ties predominantly involve international highways and airline transportation. Topological properties of airline transportation networks can explain patterns in viral disease spread worldwide \cite{Colizza2006Role,ferguson2006strategies}. At a national level, long ties pertain to mobility through major roads and trains between cities and at a regional level to commuting and local delivery services.

\section{Model}

\subsection{Small World SEIR model}
We model the spread of the disease by a small world network, in which each node of the network
is either Susceptible, Exposed, Infectious, or Recovered (an SEIR model~\cite{M, YBHB}).
In this model, nodes are individuals and we extrapolate our results to large communities
(countries, continents). 
Initially, a node is in state $S$. If it is infected, it moves through states $E$, $I$, and $R$.
While in state $I$, a node may infect each of its neighbors with a probability $r$ per time step
(fixed at one day in our simulations).
The duration
of state $E$ is the incubation period, and the duration of state $I$ is the infection period, both
of which are lognormally distributed~\cite{OSS}. The probability $r$ is our single calibration parameter and
all other parameters are fixed according to values that have been reported in the literature.
In our simulations,
the incubation period has a mean of $5$ days and the infection time has a mean of $6.5$ days,
both periods have a standard deviation of $3$ days. This is comparable to values reported in~\cite{PLRKED, Li_et_al, KRD, LiYangDang}.
Each simulation is started by infecting a single random node (patient zero).

The network in our model is the familiar Watts-Strogatz small world network, which distinguishes between short and long ties. 
For purposes of illustration a small example network with \emph{N} = 100, \emph{k} = 20, and \emph{p} = 0.05 is shown in Figure 1. 
It allows us to focus on the `stay nearby or get checked' policy.
The network is described by two parameters $k, p$, where
$k$ is the number of ties per node, which is equal for all nodes, and $p$ is the fraction of randomly selected long ties. 
In the simulations we fix \emph{N} = 10000, \emph{k} = 20, and \emph{p} = 0.1, which is in the standard
range~\cite{SWpar} and has been used to model an influenza outbreak in the Oran region of Algeria~\cite{YBHB}. Results are robust to reasonable changes in these parameters.

\begin{figure}[h!]
	\centering
    \includegraphics[width=0.35\textwidth]{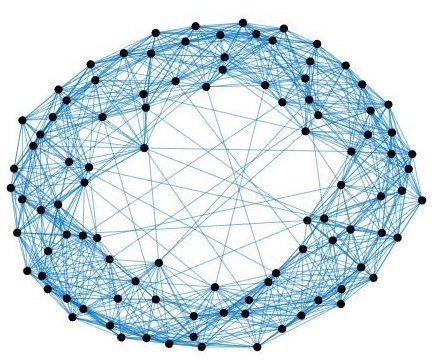}
    	\caption{\small An example of a Watts-Strogatz small-world graph with $N=100$, $k=20$ and $p=0.05$.}
\end{figure}

\subsection{Model calibration}
We calibrated our model against the number of infections in Wuhan (source: Johns Hopkins), which went into an initial lockdown
from January 23, followed by a heavy lockdown from February 10 ~\cite{LiYangDang}.
A substantial number of infections remains undetected and our simulations are based on the estimate that only ten percent of the
infected cases are officially confirmed~\cite{TenPercent}. After an exposed individual becomes infectious, it may take several days to develop symptoms, and from development of symptoms it takes an average of 5 days to diagnosis ~\cite{LiYangDang}. We use 10 days from becoming infectious to diagnosis.
We calibrated our model parameter $r$ with values $0.055$ in the period before lockdown, $0.0065$ for the initial lockdown and $0.0012$ for the severe lockdown. 
By using these values, we were able to reproduce the total number of officially confirmed cases, as demonstrated in figure~\ref{fig:dailyWuhan} below.

\begin{figure}[h!]
	%\centering
		\begin{subfigure}[b]{0.5\textwidth}
    \includegraphics[width=\textwidth,height=0.75\textwidth]{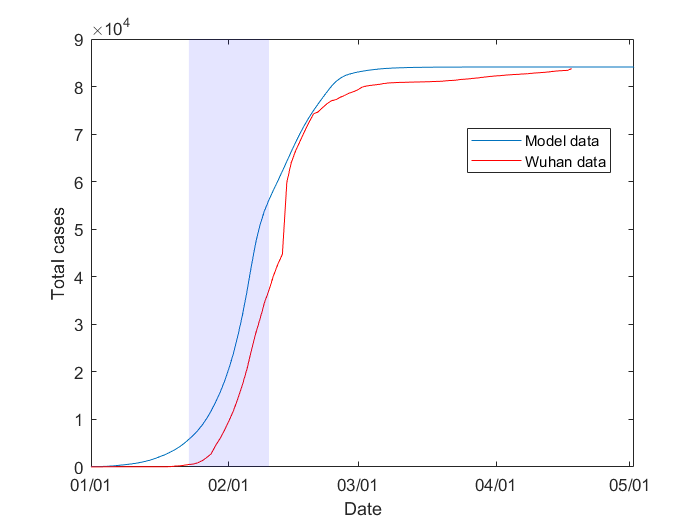}
    %\caption{Total number of reported cases in the Hubei area$}
    %\label{fig:}
  \end{subfigure}
  \begin{subfigure}[b]{0.5\textwidth}
    \includegraphics[width=\textwidth,height=0.75\textwidth]{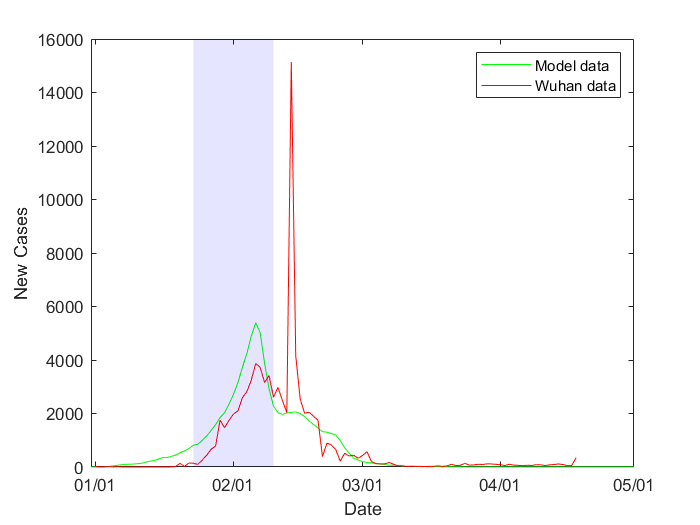}
    %\caption{Daily number of reported cases in the Hubei area}
    %\label{fig:}
  \end{subfigure}
	\caption{\small Total number of officially confirmed cases (left) and daily number of new cases in the Hubei area. The shaded part represents the initial lockdown, between Jan 23 and Feb 10. The spike in new cases on Feb 12 was due to an inclusion of previously uncounted clinically diagnosed patients. The model data are an average of 200 Monte Carlo simulations.}
	\label{fig:dailyWuhan}
\end{figure}

The reproduction number $R_0$ at the onset of the epidemic is equal to $r\cdot k\cdot t$ in our model. 
This value is high at the onset, but reduces already before the initial lockdown due to the
clustered structure of the network. We find an average $R_0=3.9$ pre-lockdown, $R_0=0.54$ during the initial lockdown
and $0.12$ during the severe lockdown. These numbers agree with the statistical analysis in~\cite{LiYangDang}.

\begin{figure}[h!]
	\centering
		\includegraphics[width=0.45\textwidth]{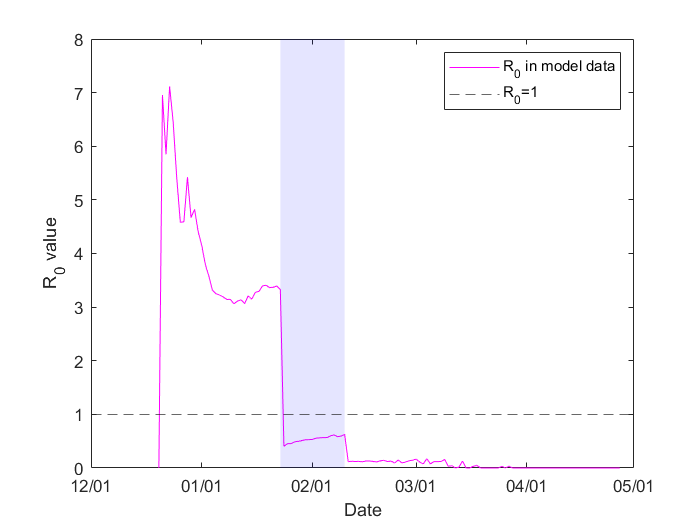}
	\caption{\small The computed reproduction number $R_0$ during the simulation reduces from an initial $7.15$ (from
	$r=0.055, k=20, t=6.5$) but drops even before the lockdown of January 23. Its average is $0.54$ during the initial lockdown
	(shaded area) and reduces even further during the severe lockdown.}
	\label{fig:R0Wuhan}
\end{figure}

We fixed $r=0.055$ for all simulations of the spread of the disease before lockdown. We calibrated $r$ after lockdown 
against the data of several European countries: Italy, Austria, Sweden, Germany.  
Italy announced
its lockdown relatively late. We used $r=0.01$ after lockdown. We find $R_0=4.0$ pre-lockdown, which is marginally higher than the $3.8$ found
in~\cite{LiYangDang}, and $R_0=0.84$ post-lockdown.
  
\begin{figure}[h!]
	\begin{subfigure}[b]{0.5\textwidth}
    \includegraphics[width=\textwidth,height=0.75\textwidth]{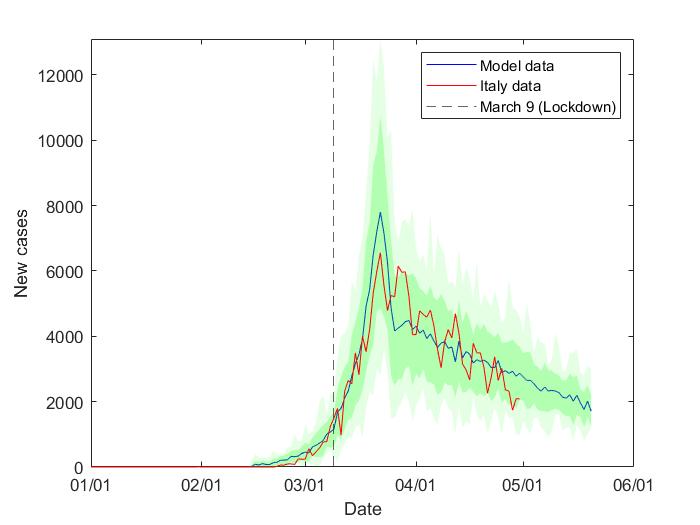}
    %\caption{Daily number of cases in Italy}
    %\label{fig:}
  \end{subfigure}
  \begin{subfigure}[b]{0.5\textwidth}
    \includegraphics[width=\textwidth,height=0.75\textwidth]{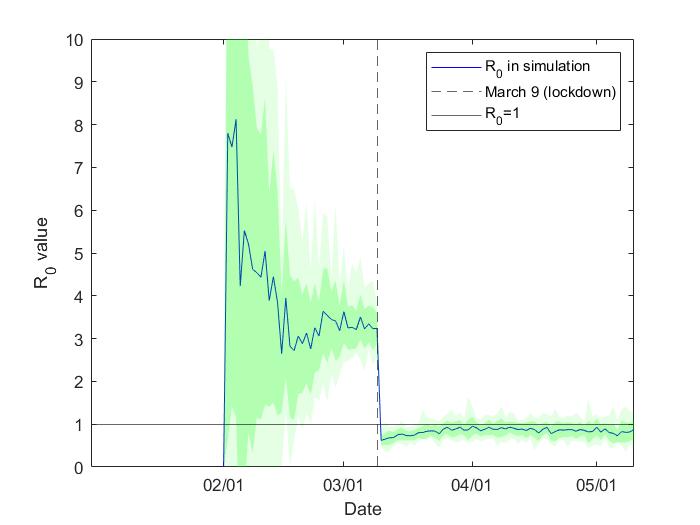}
    %\caption{$R_0$ value of our model}
    %\label{fig:}
  \end{subfigure}
	\caption{\small Daily number of officially confirmed new cases in Italy (left) with $r=0.01$ after lockdown. 
	Our simulation starts with a single infected node on January 31 and we set the initial date of the lockdown
	at March 9, when the government announced nationwide regulations.
	We plotted one standard deviation difference in dark green, and a 98 percent confidence interval in light green
	around the model data to illustrate the accuracy of our model. We find an average reproduction number $R_0=4.0$ (right) before lockdown and $R_0=0.84$ after lockdown. }
	\label{fig:dailyItaly}
\end{figure}

Austria implemented a relatively severe lockdown. We found that a parameter value of
$r=0.0055$ post-lockdown reproduces the number of confirmed cases. 

\begin{figure}[h!]
	\centering
		\includegraphics[width=0.90\textwidth]{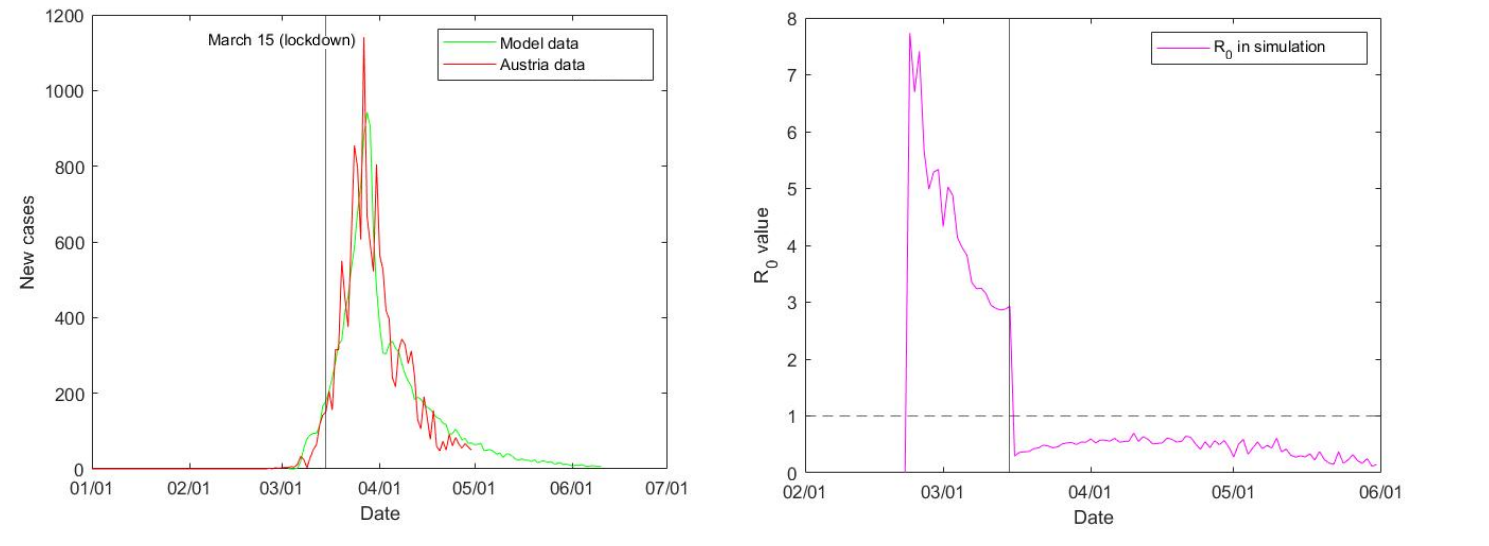}
	\caption{\small Daily number of officially confirmed new cases in Austria (left) with $r=0.0055$ after lockdown. 
	Our simulation starts with a single infected node on February 22 and a lockdown
	at March 15. We find an average reproduction number $R_0=4.2$ (right) before lockdown and $R_0=0.55$ after lockdown. }
	\label{fig:dailyAustria	}
\end{figure}

\begin{figure}[h!]
	\centering
		\includegraphics[width=0.9\textwidth]{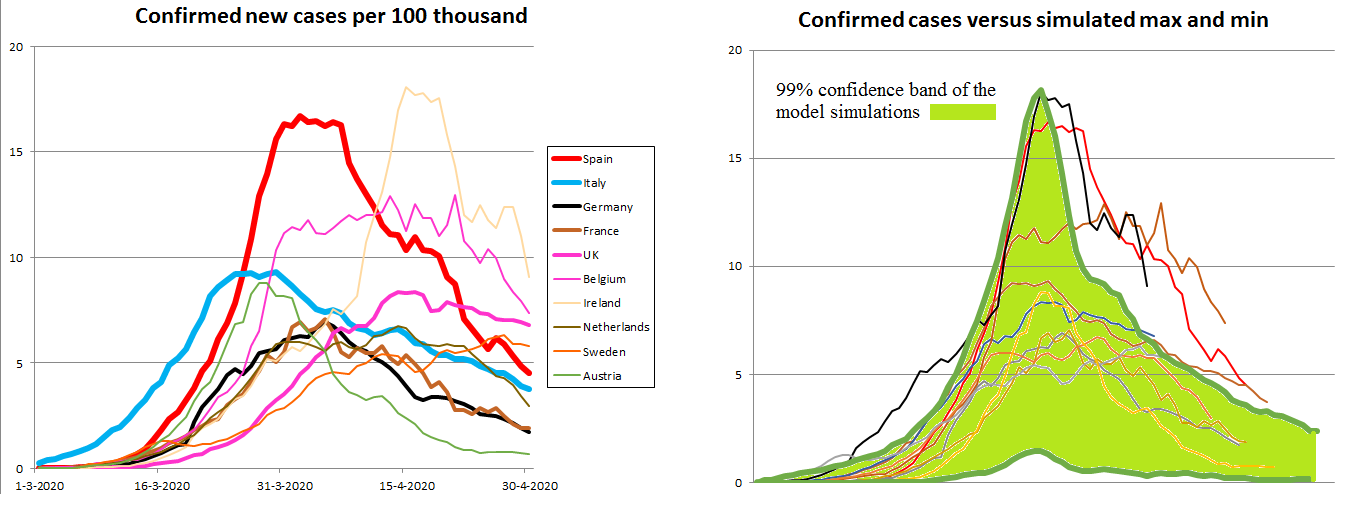}
	\caption{\small A comparison of the daily number of officially confirmed new cases in ten EU countries during
	March and April 2020. The numbers are averaged over six days. In the right hand figure, this data is adjusted
	so that the peak of the number of cases occur after 50 days. This is compared to the 99\% confidence band of
	a model computation with $r=0.055$ pre-lockdown and $r=0.0075$ during lockdown. Not all countries fit in the confidence band because their waves plateaued, while our model shows a sharp peak. The increase and decrease of the wave in our model does match the data.}
	\label{fig:dailyEU}
\end{figure}

To verify if the spread of the virus in our network model matches the observed spread of the disease, we compare 
our model to data on different scales: countries
(American states, European countries), regions (in Italy and Germany), and cities (New York City zip codes). We order
the data from largest to smallest and normalize by dividing by the largest number of cases.
We find that the spread of the disease in our network model is comparable to the data.

\begin{figure}[h!]
	\centering
		\includegraphics[width=0.55\textwidth]{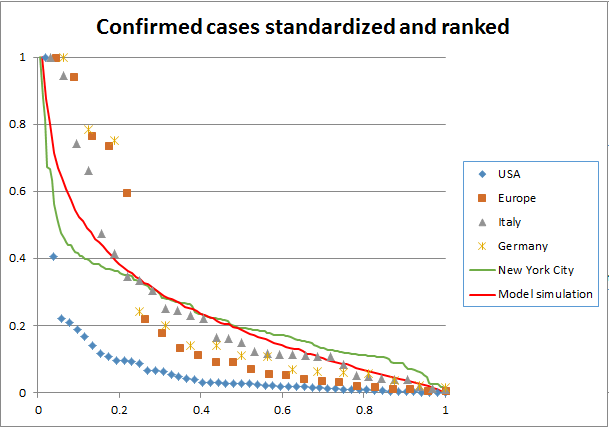}
	\caption{\small Number of confirmed cases on three different scales: USA and Europe -- Regione and Bundesl\"ander -- NYC zip codes.
	The data is ranked from largest to smallest and normalized. On each scale, the spread of the disease displays a similar exponential
	decay. The figure also shows the spatial distribution of Covid-19 spread in our model after 60 days, 25 days into lockdown. To this end we arbitrarily divided the ring lattice of $N$ = 10,000 nodes into 100 regions of 100 nodes each. The spatial distribution in
		our network is comparable to the data of Europe, Germany, Italy, and New York City. The spread of the disease in the USA is more
		concentrated, with an exceptional number
		of cases in New York and New Jersey.
	}
	\label{fig:rankplot}
\end{figure}	

\begin{figure}[h!]
\begin{tabular}{|c|l|}
\hline $r$&intepretation\\
\hline
$0.001$& {\small Complete lockdown as in Wuhan after February 10th. Fully controlled society.}\\
$0.005$& {\small Severe lockdown as in Austria. Strong restrictions on travel, shopping, gatherings.}\\
$0.01$& {\small Moderate regulations and social distancing. As in many EU countries after March 15.}\\
$0.02$& {\small Requested social distancing, but no regulations.}\\
%As in Sweden after March 10. %no, sweden has some regulation, and r=0.125}\\
$0.055$& {\small Pre-lockdown situation. No social distancing.}\\
\hline
\end{tabular}
\end{figure}

\section{Impact of a stay-nearby-or-get-checked policy}

\subsection{Peak reduction}

As countries get out of lockdown, restrictions are lifted, and the basic reproduction number will increase, possibly inducing a second wave. 
Our proposed policy is to control the long ties to flatten the second wave. To simulate this policy, we use the parameters of the calibrated model. During an initial lockdown of 785 days, spreading risk is reduced to $r=0.0075$. After lockdown, the parameter is increased to $r=0.02$ (requested social distancing, but no regulations), but some fraction of edges is checked, preventing propagation along those edges. Figure 8 contrasts a scenario in which random edges are checked (panel A) with one in which only long edges are checked (panel B). Results show that targeting long edges is efficient. Without intervention a second wave occurs with a peak several times higher than the first. By targeting long ties, checking 7.5\% of ties is sufficient to bring the second peak below the first peak, thereby also delaying its occurrence. If instead random ties are chosen the second wave peak remains well above the first. 

\begin{figure}[h!]
\begin{adjustwidth}{0em}{0em}
		\begin{subfigure}[b]{0.5\textwidth}
    \includegraphics[width=\textwidth,height=0.75\textwidth]{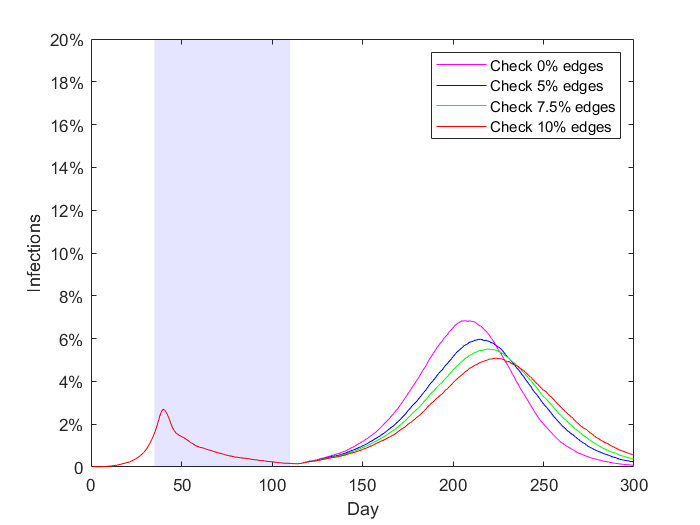}
    \caption{Check random edges}
    %\label{fig:}
  \end{subfigure}
  \begin{subfigure}[b]{0.5\textwidth}
    \includegraphics[width=\textwidth,height=0.75\textwidth]{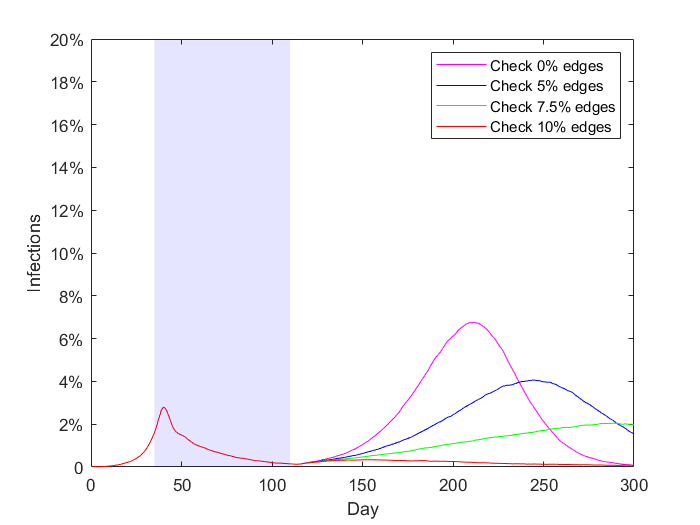}
    \caption{Check long edges}
    %\label{fig:}
  \end{subfigure}
  \end{adjustwidth}
	\caption{\small Peak reduction with baseline $r$ = 0.02 and some edges checked}
	\label{fig:budget}
\end{figure}

In figure 9 we contrast three post-lockdown strategies for controlling a second wave. Here we use $r_{short}$ to denote transmission probabilities on all short ties and and $r_{long}$ transmission probabilities on long ties. The first strategy is to not differentiate long ties from short ties (panel A). Results show that at $r_{long}$ = $r_{short}$ = 0.02 (requested social distancing without further regulations) the second wave peaks much higher than the first, and that further regulation is needed to control it.

The second strategy is to open up society back to what it was before the pandemic ($r_{short}$ = 0.055), with no social distancing, and then target the 10\% of long ties for checking. Panel B shows various levels of effectiveness in intervening on long ties, varying $r_{long}$. Panel B shows that this strategy only works if disease propagation in long ties can be fully prevented. In panel C, society is opened back up but social distancing is requested at an $r$ = 0.02. Now, imperfect checking of long ties  $r_{long}$ = 0.005 can also accomplish a reduction of the second peak below the first.

\begin{figure}[hbtp!]
\begin{adjustwidth}{0em}{0em}
\centering
  \begin{subfigure}[b]{0.5\textwidth}
    \includegraphics[width=\textwidth,height=0.75\textwidth]{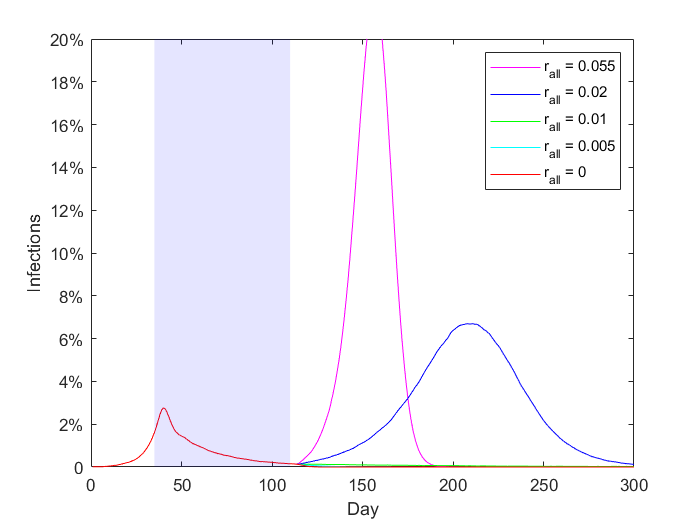}
    \caption{Vary $r_{short}$ and $r_{long}$ together}
    %\label{fig:vary r all}
  \end{subfigure}
  %
  %\begin{subfigure}[b]{0.5\textwidth}
  %  \includegraphics[width=\textwidth,height=0.75\textwidth]{version2/Vary_r_short_50_sims.png}
  %  \caption{Vary $r_{short}$, keep $r_{long}=0.055$ }
  %  %\label{fig:vary r short long=0055}
  %\end{subfigure}
\end{adjustwidth}
\begin{adjustwidth}{0em}{0em}
  \begin{subfigure}[b]{0.5\textwidth}
    \includegraphics[width=\textwidth,height=0.75\textwidth]{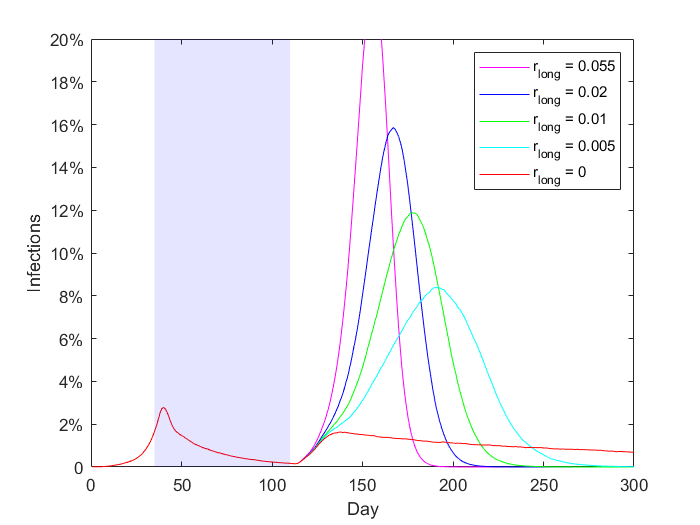}
    \caption{Vary $r_{long}$, keep $r_{short}=0.055$}
    %\label{fig:vary r long short=0055}
  \end{subfigure}
  \begin{subfigure}[b]{0.5\textwidth}
    \includegraphics[width=\textwidth,height=0.75\textwidth]{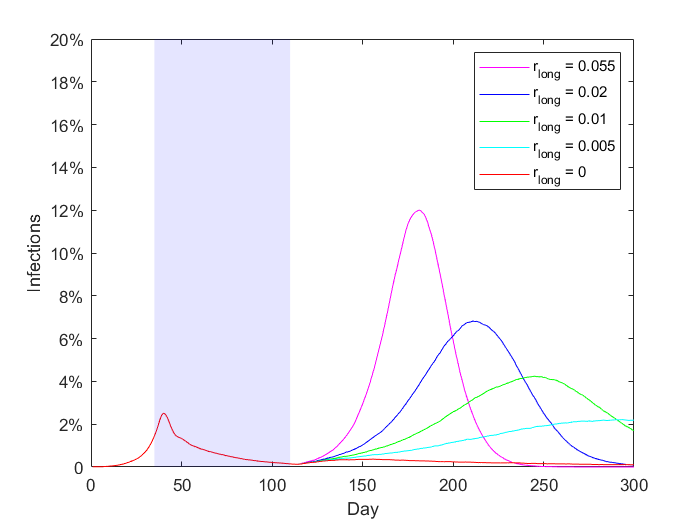}
    \caption{Vary $r_{long}$, keep $r_{short}=0.02$}
    %\label{fig:vary r long short=002}
  \end{subfigure}
  \end{adjustwidth}
  \caption{\small Three policy approaches to opening up a lockdown of 75 days, with varying post-lockdown levels of $r$: (a) policy does not differentiate long and short ties, (b) policy targets long ties while all other restrictions are lifted, and (c) policy targets long ties while social distancing is encouraged.}
\end{figure}

%%% This figure is now obsolete, replaced by figure of 4. %%%%%
% \begin{figure}[h!]
% 	\centering
%     \includegraphics[width=4in]{2ndWave005}
%     	\caption{Effect of varying long-range transmission probability $r_{long}$ after day 100 on second wave peak. $r_{long}$ and $r_{short}$ start at 0.1 on day 1, and switch to 0.01 on day 15. On day 100 $r_{short}$ switches back to 0.05.}
% \end{figure}

\subsection{Spatial concentration}

Post-lockdown flare-ups are more easily controlled with geographically focused efforts when they remain local longer. Economic and social costs of control measures are then also lower. We study the spatial concentration of Covid-19 outbreaks by measuring the number of components of the subgraph of infected nodes and short edges. Figure 9 compares the spread of the virus for the scenario where $r_{long} = r_{short}$ = 0.055 with the alternative scenario where long-range transmission is maximally repressed, $r_{long} = 0$. The latter scenario is characterized by a smaller number of components during the second wave. The spread of the virus with $r_{long}=0$ is contained in one region.

\begin{figure}[h!]
\begin{adjustwidth}{0em}{0em}
	\centering
    \includegraphics[width=\textwidth]{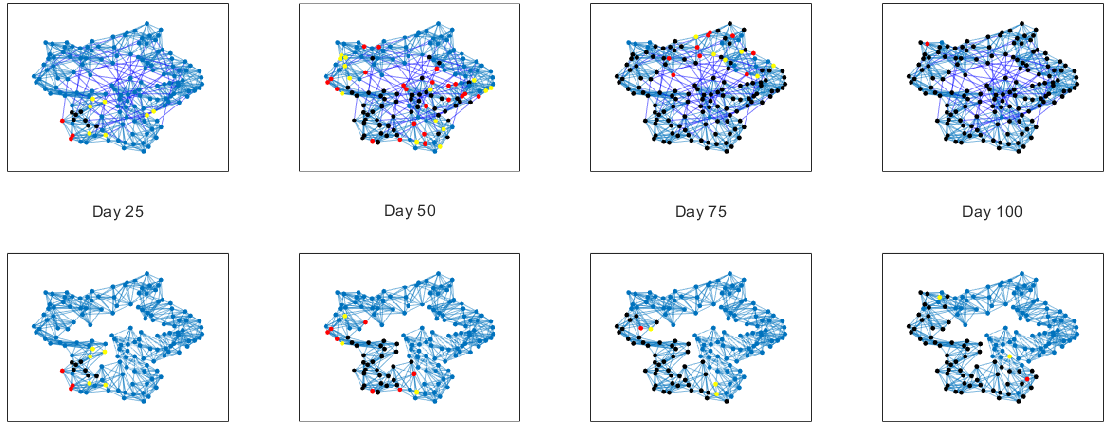}
    \end{adjustwidth}
    	\caption{Effect of shutting down long-range transmission. From left to right graphs show disease propagation at 25 day intervals in a small-world network with $N=150$, $k=10$ and $p=0.1$. All graphs have $r_{short}=0.055$. The top graphs have $r_{long}=0.055$, while the bottom graphs have $r_{long}=0$. Node color indicates SEIR states, blue = susceptible, yellow = exposed, red = infectious, black = resistant.}
    	%\label{fig:animation}
\end{figure}

\section{Discussion and Policy}
In this paper we have through model simulation explored spatially differentiating policies in which interventions target nonlocal spread of Covid-19. Our results show that reductions of long-distance transmission are highly efficient for curbing the spread of Covid-19. The close monitoring and checking of long-distance ties allows overall policy to be more permissible and still control a second wave. Because flare-ups remain local longer, interventions can be of limited geographical scope and thus less costly and invasive.

What policies could constrain long-range transmission? Medical testing and mobility-tracking apps may be targeted specifically at transport, travel, and delivery. Perhaps medical testing and / or mobility tracking should be encouraged or required for flight, use of highways, trains, regional bus lines and for individuals with jobs in the transport and delivery sector. Self-isolation after exposure of such individuals may perhaps be more stringently enforced. What helps is that long-range ties are relatively sparse, so resources may be focused on a limited number of individuals or activities. That said, our results show that effects are particularly strong when transmission through long-range ties is not just reduced but largely eliminated. This concords with studies showing that international traffic constraints are particularly effective when severe \cite{ferguson2006strategies}. The logistical, technological and ethical challenges of geographic targeting in location tracking, testing, and police enforcement require further interdisciplinary study.

\bibliographystyle{unsrt}
\bibliography{nearbyorchecked} 

\end{document}